\documentclass[preprint,showpacs,preprintnumbers,amsmath,amssymb]{revtex4}
\usepackage{booktabs}
\usepackage{mathrsfs}
\usepackage{epsfig}
\usepackage{graphicx}
\usepackage{dcolumn}
\usepackage{bm}
\usepackage{amsmath}
\usepackage{slashed}
\usepackage{multirow}

\let\jnfont=\rm
\def\NPB#1,{{\jnfont Nucl.\ Phys.\ B }{\bf #1},}
\def\PLB#1,{{\jnfont Phys.\ Lett.\ B }{\bf #1},}
\def\EPJC#1,{{\jnfont Eur.\ Phys.\ Jour.\ C }{\bf #1},}
\def\PRD#1,{{\jnfont Phys.\ Rev.\ D }{\bf #1},}
\def\PRL#1,{{\jnfont Phys.\ Rev.\ Lett.\ }{\bf #1},}
\def\MPLA#1,{{\jnfont Mod.\ Phys.\ Lett.\ A }{\bf #1},}
\def\JPG#1,{{\jnfont J.\ Phys.\ G}{\bf #1},}
\def\CTP#1,{{\jnfont Commun.\ Theor.\ Phys.\ }{\bf #1},}
\def\ZPC#1,{{\jnfont Z.\ Phys.\ C }{\bf #1},}
\def\JHEP#1,{{\jnfont JHEP \ }{\bf #1},}
\def\Rv{\not{\hbox{\kern-1pt $R$}}}
\def\p{\not{\hbox{\kern-3pt $p$}}}

\begin{document}

\title{Enhancing $thj$ Production from Top-Higgs FCNC Couplings}

\author{ Lei Wu }
\affiliation{ ARC Centre of Excellence for Particle Physics at the Terascale, School of Physics, The University of Sydney, NSW 2006, Australia}%

\date{\today}

\begin{abstract}
In this paper, we study the single top and Higgs associated production $pp \to thj$ in the presence of top-Higgs FCNC couplings($\kappa_{tqh}, q=u,c$) at the LHC. Under the current constraints, we find that the cross section of $pp \to thj$ can be sizably enhanced in comparison with the SM predictions at 8 and 14 TeV LHC. We also find that the full cross section of $pp \to thj$ with $\kappa_{tch}$ is larger than the resonant cross section of $pp \to t\bar{t} \to thj$ by a factor 1.16 at 8 TeV LHC and 1.12 at 14 TeV LHC, respectively. We further explore the observability of top-Higgs FCNC couplings through $pp \to t(\to b\ell^{+} \nu_{\ell}) h( \to \gamma\gamma) j$ and find that the branching ratios $Br(t\to qh)$, $Br(t \to uh)$ and $Br(t \to ch)$ can be respectively probed to $0.12\%,~0.23\%$ and $~0.26\%$ at $3\sigma$ sensitivity at 14 TeV LHC with ${\cal L} =3000$ fb$^{-1}$.
\end{abstract}
\maketitle

\section{INTRODUCTION}
The discovery of the Higgs boson at the LHC is a great triumph of the Standard Model(SM) and marks a new era in the particle physics \cite{atlas,cms}. Given the  large uncertainties of the current Higgs data, there remains a plenty of room for new physics in Higgs sector \cite{Higgs-review}. So the precise measurement of the Higgs boson's properties will be a dominant task at the LHC in the next decades.

Concerning the probe of new physics through the Higgs boson, the Yukawa couplings can play the important role since they are sensitive to new flavor dynamics beyond the SM. In particular, top quark, as the heaviest SM fermion, owns the strongest Yukawa coupling and has the preference to reveal the new interactions at the electroweak scale \cite{top-review}. One of the interesting things is that the top quark is just heavier than the observed Higgs boson, which makes the top quark flavor changing neutral current(FCNC) processes $t \to h q~(q=u,c)$ be accessible in kinematics. In the SM, these top quark FCNC transitions are extremely suppressed by the G.I.M. mechanism \cite{tcvh-sm}. But they can be greatly enhanced by the extended flavor structures in many new physics models, for example the minimal supersymmetric model (MSSM) with/without R-parity \cite{rpc,rpv}, two-Higgs-doublet model(2HDM) type-III \cite{2hdm-1,2hdm-2}, and the other miscellaneous models \cite{other1,other2,other3}. So the study of top-Higgs FCNC interactions is a common interest of the theory and experiment communities \cite{top-fcnc-th,top-fcnc-exp,atlas-fcnc,cms-fcnc}. However, up to now, the null results of the searches for $t \to qh$ at the LHC give the strong limits on the top-Higgs FCNC couplings. Among them, the most stringent constraint $Br(t \to hc) < 0.56\%$ at $95\%$ C.L. was reported by the CMS collaboration from a combination of the multilepton channel and the diphoton plus lepton channel \cite{cms-fcnc}. Except for the widely studied $t \to qh$ decays, the importance of the single top+Higgs production $pp \to th$ in probing the top-Higgs FCNC couplings has been also emphasized in the recent theoretical studies \cite{th-1,th-2,th-3,th-4,th-5,th-6}.

In this paper, we investigate the top-Higgs FCNC interactions through $pp \to thj$ with the sequent decays $t \to b \ell^+ \nu$ and $h \to \gamma\gamma$ at the LHC. In the SM, the process $pp \to thj$ can only be induced by the weak charged current interaction and has a relative small cross section, which is about 18 (88) fb at 8 (14) TeV LHC. However, such a process is found to be very sensitive to modifications of the Higgs couplings \cite{tim,maltoni_thj1,maltoni_thj2,biswas_thj,ellis_thj,englert,lee_thj,thj}. In particular, the top-Higgs FCNC couplings $tqh(q=u,c)$ can sizably enhance $thj$ cross section due to the contributions from the strong interaction processes. To be specific, there are mainly three new kinds of processes that can contribute to the production of $thj$ at the LHC: (1) gluon fusion $gg \to thj$, it is the dominant contribution, as shown in Fig.\ref{gg}, where $hj$ can be produced not only by an on-shell top quark but also by an off-shell top quark via the new flavor changing couplings $tqh$; (2) $qg$ fusion $qg \to thj$, as shown in Fig.2, which is the sub-leading contribution. However, such a process is affected by the initial Parton Distributions Functions(PDFs). So one can use this feature to disentangle the FCNC couplings of the top quark with light quarks; (3) $q\bar{q}$ annihilation $q\bar{q} \to thj$, which is similar to the $s$-channel of the process $gg \to thj$ but with $q\bar{q}$ instead of $gg$ in the initial states. The contribution of this process is relatively smaller than (1) and (2) because of the suppression of color factor and PDFs; Based on the above considerations, it is worthwhile to perform a complete calculation of $pp \to thj$ in the presence of the top-Higgs FCNC couplings by including the contributions (1)-(3), and explore its sensitivity to probe the top-Higgs FCNC couplings at the LHC.

This paper is arranged as follows. In Sec. II, we set up the notations and briefly describe the top-Higgs FCNC interactions. In Sec. III, we discuss the observability of the top-Higgs FCNC couplings through the process $pp \to thj$ at 14 TeV LHC. Finally, a summary is given in Sec. IV.

\section{top-Higgs FCNC interactions}
\begin{figure}[ht]
\centering
\includegraphics[width=4in,height=2in]{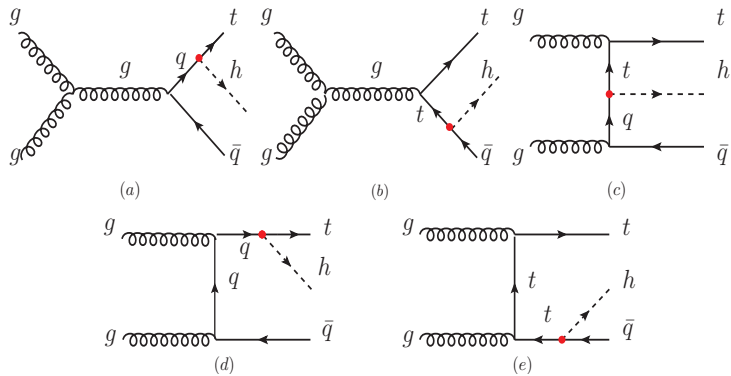}\vspace{-0.5cm}
\caption{Example Feynman diagrams for the partonic process $gg \to th\bar{q}$ at the LHC through flavor violating top-Higgs interactions in Eq.(\ref{tqh})(marked with red dots). Here $q=u,c$.}
\label{gg}
\end{figure}
\begin{figure}[ht]
\centering
\includegraphics[width=4in,height=1in]{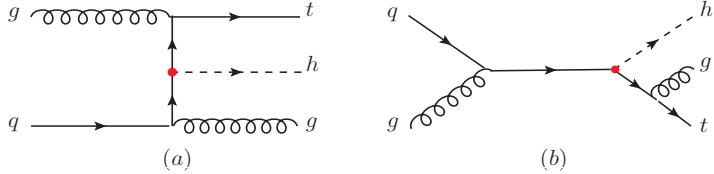}\vspace{-0.5cm}
\caption{Example Feynman diagrams for the partonic process $qg \to thg$ at the LHC through flavor violating top-Higgs interactions in Eq.(\ref{tqh})(marked with red dots). Here $q=u,c$.}
\label{qg}
\end{figure}
A general effective Lagrangian describing the top-Higgs FCNC interaction can be written as
\begin{equation}
-{\cal L}_{tqh}= \kappa^L_{tqh}\bar{t}_{L}q_{R}h+\kappa^R_{tqh}\bar{t}_{R}q_{L}h+h.c..
\label{tqh}
\end{equation}
where $h$ is the SM Higgs boson, and the real parameter $\kappa^{L,R}_{tqh}$ denote the left-handed and right-handed FCNC couplings of the Higgs boson to the light up-type quarks $q=u,c$. We plot example Feynman diagrams in Fig.\ref{gg} and Fig.\ref{qg} for the partonic process $gg \to th\bar{q}$ and $qg \to thg$, respectively. Some diagrams of the process $u\bar{u}/d\bar{d} \to th\bar{q}$ can be obtained by replacing the initial gluons with $u\bar{u}/d\bar{d}$ in the $s$-channel in Fig.\ref{gg}. By neglecting the light quark masses and assuming the dominant top decay width $t \to bW$, the Leading Order(LO) branching ratio of $t \to qh$ can be approximately given by,
\begin{equation}
Br(t \to qh) = \frac{{\kappa^{L}_{tqh}}^2+{\kappa^{R}_{tqh}}^2}{2\sqrt{2} m^2_t G_F}\frac{(1-x^2_h)^2}{(1-x^2_W)^2 (1+2x^2_W)}.
\end{equation}
where $G_F$ is the Fermi constant, $x_W=m_W/m_t$ and $x_h=m_h/m_t$. The NLO QCD correction to $Br(t \to qh)$ is estimated as $10\%$ according to the results of high order corrections to $t\to bW$ \cite{twb} and $t \to ch$ \cite{nlo}. In some specific models, the left-handed coupling $\kappa^{L}_{tqh}$ is not expected to be large because its relation with the CKM mixing parameter. Also, $\sqrt{{\kappa^{L}_{tqh}}^2+{\kappa^{R}_{tqh}}^2}$ can be constrained by the low energy observables, such as $B^0-\overline{B^0}$ mixing \cite{2hdm-2,b0b0}. However, we do not consider these indirect constraints in our study since they are model-dependent and their relevance strongly depends on the assumptions made for the generation of the quark flavor structures \cite{parameterization}. On the other hand, the CMS collaboration reported a model-independent bound $\sqrt{{\kappa^{L}_{tqh}}^2+{\kappa^{R}_{tqh}}^2} < 0.14$ at $95\%$ C.L. from the combined result of multilepton and diphoton in $t\bar{t}$ production \cite{cms-fcnc}, which indicates $|\kappa^{L,R}_{tqh}|$ should be less than 0.14. In our work, we assume $\kappa^{L}_{tqh}=\kappa^{R}_{tqh}=\kappa_{tqh}$ and require $\kappa_{tqh} \leq 0.1$ to satisfy the direct constraint from the CMS result.

\section{Numerical Calculations and discussions}

We implement the top-Higgs FCNC interactions by using the package \textsf{FeynRules} \cite{feynrules} and calcualte the LO cross section of $pp \to thj$ with \textsf{MadGraph5} \cite{mad5}. We use CTEQ6L as the parton distribution function(PDF) \cite{cteq} and set the renormalization scale $\mu_R$ and factorization scale $\mu_F$ to be $\mu_R=\mu_F=(m_t + m_h)/2$. The SM input parameters are taken as follows \cite{pdg}:
\begin{align}
m_t = 173.07{\rm ~GeV},\quad &m_{Z} =91.1876 {\rm ~GeV}, \quad \alpha(m_Z) = 1/127.9, \\ \nonumber
\sin^{2}\theta_W = 0.231,\quad &m_h =125 {\rm ~GeV}, \quad \alpha_{s}(m_Z)=0.1185.
\end{align}

\begin{figure}[h]
\centering
\includegraphics[width=6in,height=4in]{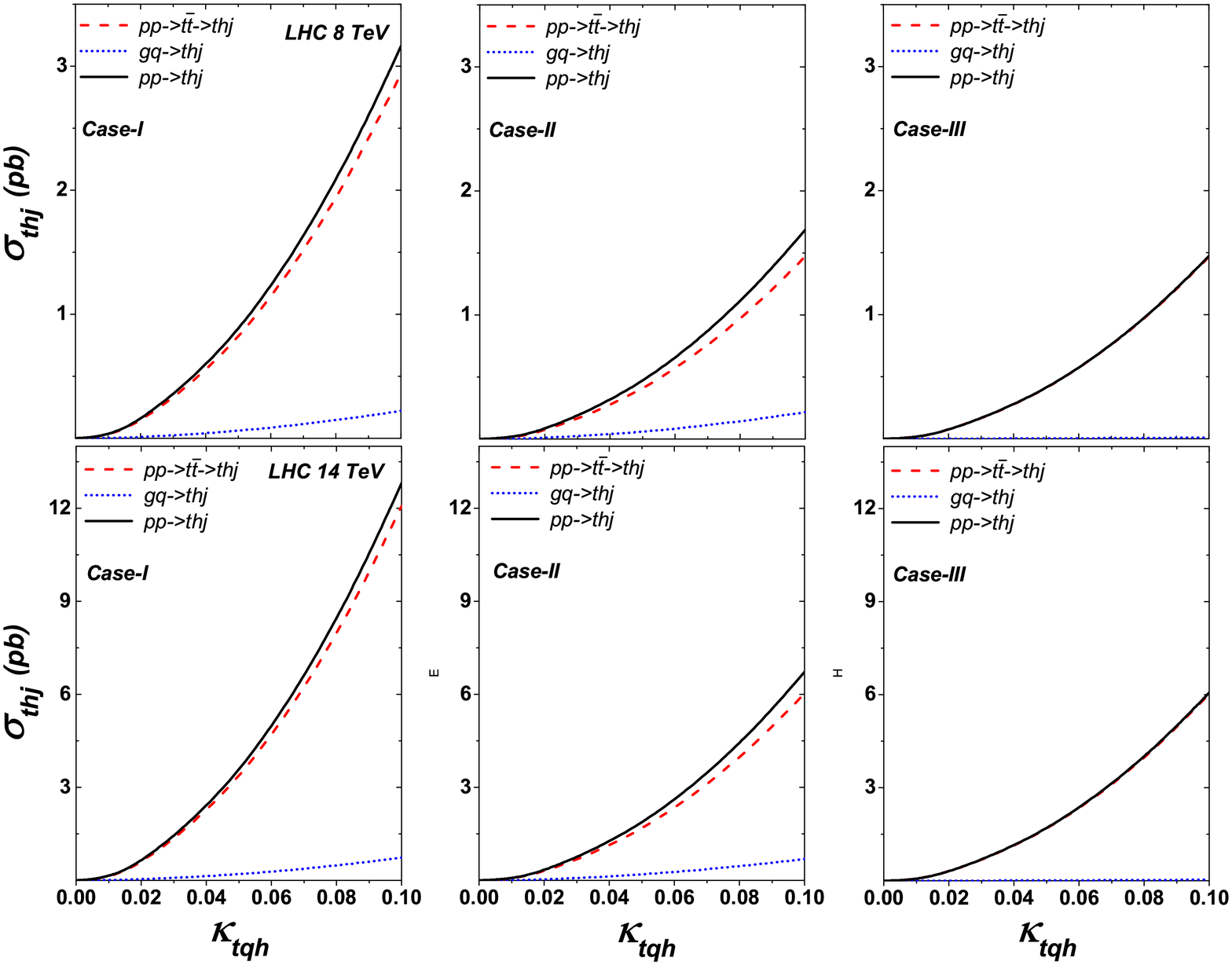}\vspace{-0.5cm}
\caption{The dependence of the cross sections $\sigma_{thj}$ at 8 and 14 TeV LHC on the top-Higgs FCNC couplings $\kappa_{tqh}$ for case $(I)-(III)$. The conjugate processes have been included in the calculations.}
\label{constraint}
\end{figure}
In Fig.\ref{constraint}, we show the dependence of the cross sections $\sigma_{thj}$ on the top-Higgs FCNC couplings $\kappa_{tqh}$ at 8 and 14 TeV LHC respectively for three different cases: $(I)~\kappa_{tqh}=\kappa_{tuh}=\kappa_{tch}$, $(II)~\kappa_{tqh}=\kappa_{tuh}, \kappa_{tch}=0$ and $(III)~\kappa_{tqh}=\kappa_{tch}, \kappa_{tuh}=0$. For the three cases, the main contribution to $pp \to thj$ is from the resonant production $pp \to t\bar{t} \to thj$. The non-resonant contributions are dominated by the process $gq \to thj$. To be specific, we can have the following observations:
\begin{itemize}

  \item  Case-$(I)$: When $\kappa_{tqh}=0.1$, the total cross section of $pp \to thj$ at 8 and 14 TeV LHC can be respectively enhanced up to nearly 176 and 145 times the SM predictions \cite{maltoni_thj2}. For the smaller values of $\kappa_{tqh}$, the cross section will decrease and become comparable with the SM prediction when $\kappa_{tqh}\sim 0.008$.  Here it should be mentioned that although the CMS collaboration has performed a search for $thj$ event at $\sqrt{s}=8$ TeV and given a 95\% upper limit on the $thj$ cross section $\sigma_{thj} < 2.24 ~ \rm pb$, this bound is not suitable for our case because a forward jet with $|\eta|>1.0$ is required in the experimental analysis. We can also see that the full cross section of $pp \to thj$ is 1.08 (1.06) times larger than the one of $pp \to t\bar{t} \to thj$ at 8 (14) TeV LHC due to the contributions of the non-resonant productions of $hj$.

  \item Case-$(II)$ and $(III)$: For the same values of $\kappa_{tuh}$ and $\kappa_{tch}$, the cross section of $pp \to thj$ in case-$(II)$ is much larger than that in case-$(III)$, since the up-quark has the larger PDF than the charm-quark. This feature allows us to separately probe the couplings between $\kappa_{tuh}$ and $\kappa_{tch}$ at the LHC. So, in general, for a given collider energy and luminosity, we can expect the sensitivity to the coupling $\kappa_{tuh}$ will be better than $\kappa_{tch}$. It should be also mentioned that the dominant contribution to $pp \to thj$ in case-$(II)$ and case-$(III)$ still come from $gg$-fusion process. The main difference between case-$(II)$ and case-$(III)$ lies in the contribution of $qg \to thj$ process. This makes the complete cross section of $thj$ almost same as that of $t\bar{t} \to thj$ in case-$(III)$ because of the small portion of $c$ quark in the proton. To be specific, when $\sqrt{s}= 8 (14)$ TeV and $\kappa_{tqh}=0.1$, $\sigma_{pp \to thj}$ is about 1.16(1.12) and 1.006(1.005) times larger than $\sigma_{pp \to t\bar{t} \to thj}$ in case-$(II)$ and -$(III)$, respectively.

  \item Case-$(I)$ and $(II)$: We also find that the impact of $qg \to thj$ on increasing the cross section of $thj$ production in case-$(I)$ is smaller than that in case-$(II)$. The reason is that the main production mode $gg \to thj$ in case-$(I)$ includes both of $gg \to th\bar{u}$ and $gg \to th\bar{c}$, while in case-$(II)$ only the former process can contribute to $gg \to thj$ production. On the other hand, the cross section of $qg \to thj$ is almost same in case-$(I)$ and $(II)$ since it is dominated by the subprocess $ug \to thg$.

\end{itemize}

In the following calculations, we perform the Monte Carlo simulation and explore the sensitivity of 14 TeV LHC to the top-Higgs FCNC couplings through the channel,
\begin{equation}\label{signal}
pp \to t(\to b\ell^{+} \nu_{\ell}) h( \to \gamma\gamma) j,
\end{equation}
which is characterized by two photons appearing as a narrow resonance centered around the Higgs boson mass. So the SM backgrounds to the Eq.(\ref{signal}) include two parts: the resonant and the non-resonant backgrounds. For the former, they mainly come from the processes that have a Higgs boson decaying to diphoton in the final states, such as $Whjj$, $Zhjj$ and $t\bar{t}h$ productions. The additional jets in the $Whjj/Zhjj$ events come from the initial or final state radiations. The cross sections of the resonant backgrounds are normalized to their NLO values; For the latter, the main background processes contain the diphoton events produced in association with top quarks, such as $tj\gamma\gamma$ and $t\bar{t}\gamma\gamma$. The $Wjj\gamma\gamma$ production can also mimic the signal when one light jet is mistagged as a $b$ jet.

We generate signal and backgrounds events with MadGraph5 and perform the parton shower and the fast detector simulations with \textsf{PYTHIA} \cite{pythia} and \textsf{Delphes} \cite{delphes}. When generating the parton level events, we assume $\mu_R=\mu_F$ to be the default event-by-event value. We cluster the jets by setting the anti-$k_t$ algorithm with a cone radius $\Delta R=0.5$ \cite{anti-kt}. The $b$-jet tagging efficiency($\epsilon_b$) is formulated as a function of the transverse momentum and rapidity of the jets \cite{cms-b}. The mis-tag of QCD jets is assumed to be the default value as in \textsf{Delphes}. In our simulation, we generate 100k events for the signals and backgrounds respectively.

\begin{figure}[ht]
\centering
\includegraphics[width=3in,height=2.5in]{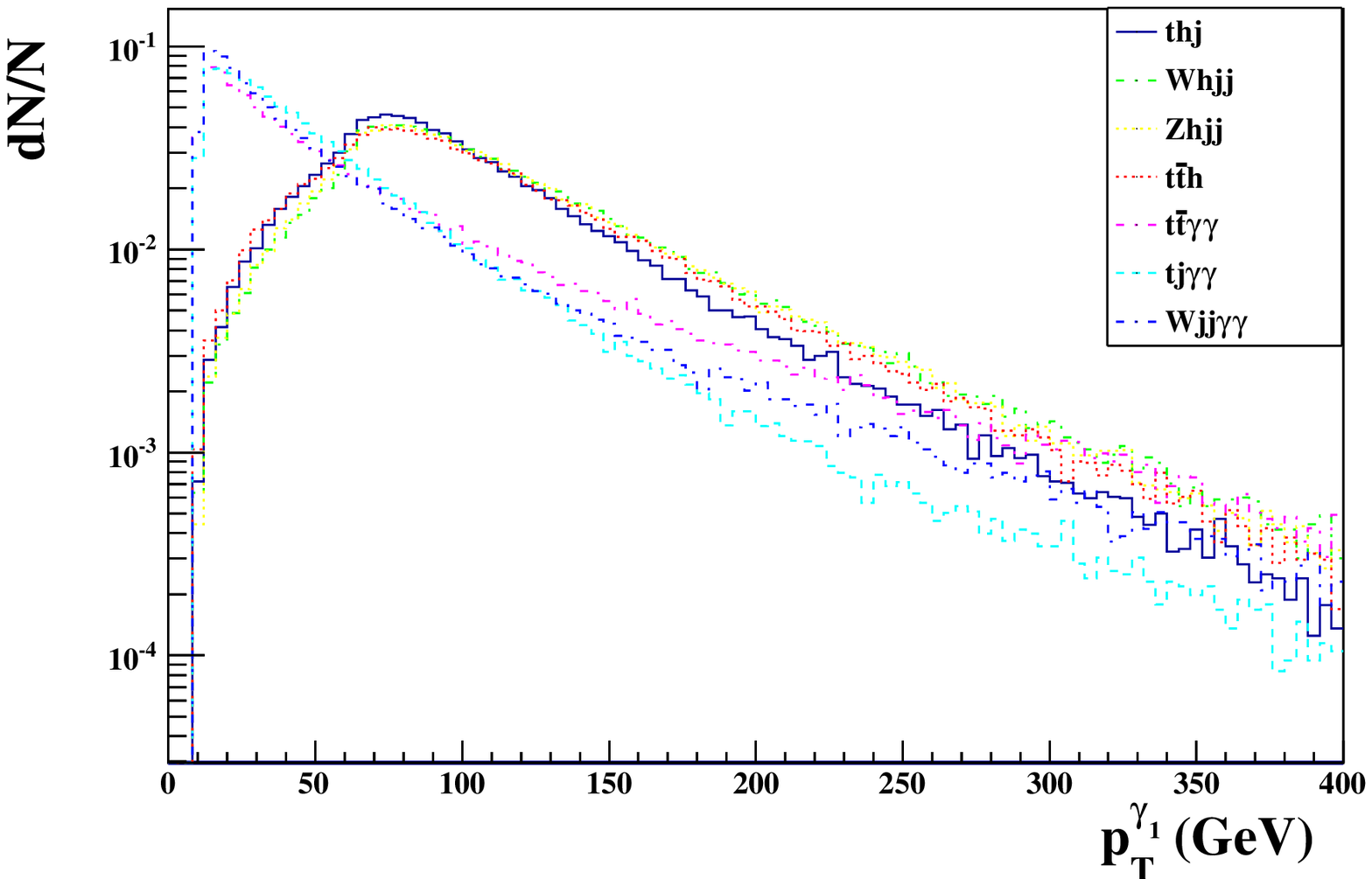}\hspace{1cm}\hspace{-1cm}
\includegraphics[width=3in,height=2.5in]{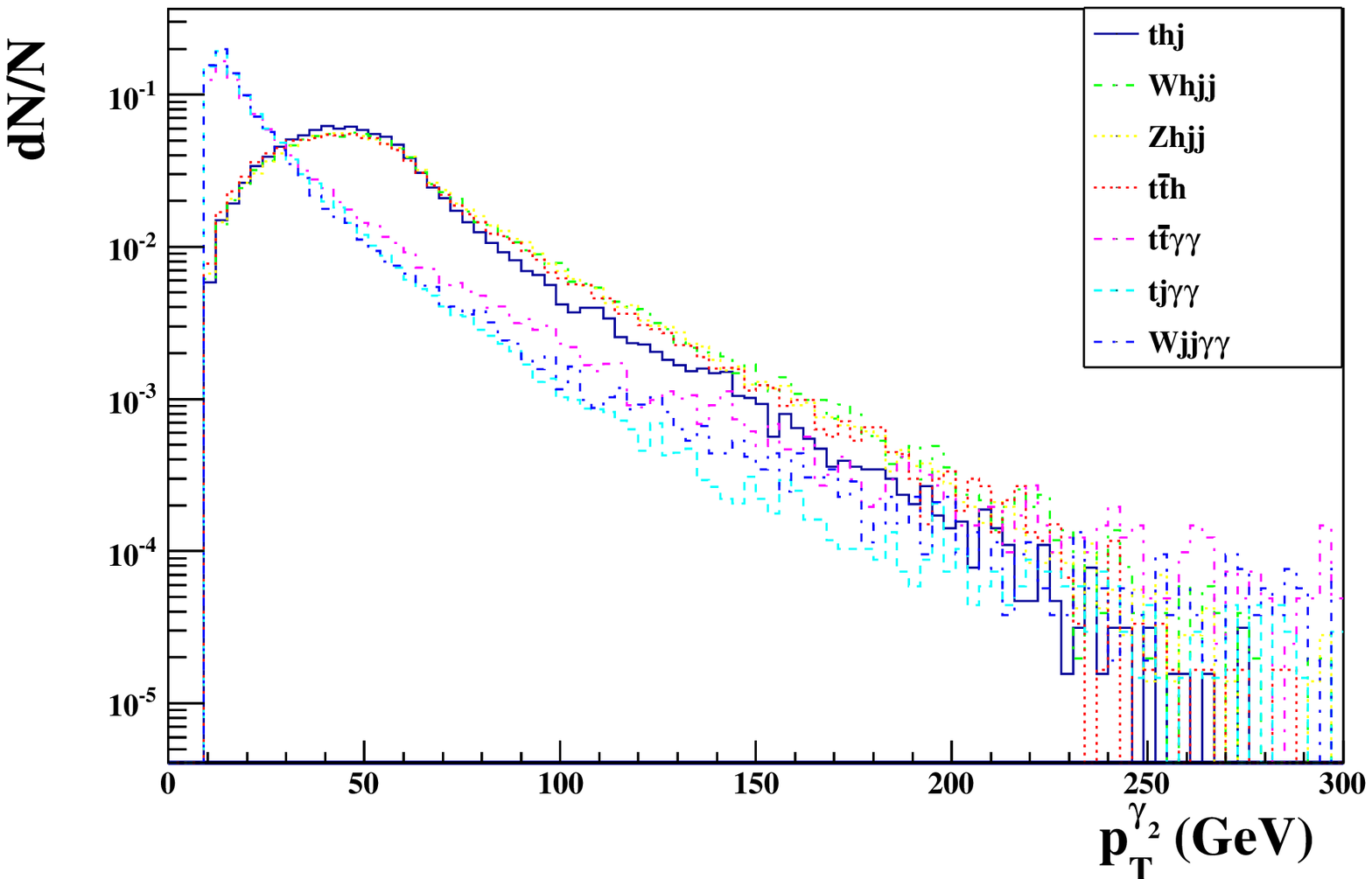}\vspace{-0.5cm}
\caption{Normalized transverse momentum distributions of two photons in the signals and backgrounds at 14 TeV LHC.}
\label{photon}
\end{figure}
In Fig.\ref{photon}, we show the transverse momentum distributions of two photons in the signal with $\kappa_{tqh}=0.1$ and backgrounds at 14 TeV LHC. Since the two photons in the signal and the resonant backgrounds come from the Higgs boson, they have peaks around $m_h/2$ and possess the harder $p_T$ spectrum than those in the non-resonant backgrounds. According to Fig.\ref{photon}, we can impose the cuts $p^{\gamma_1}_{T}>50$ GeV and $p^{\gamma_2}_{T}>25$ GeV to suppress the non-resonant backgrounds.

\begin{figure}[ht]
\centering
\includegraphics[width=3.5in,height=2.8in]{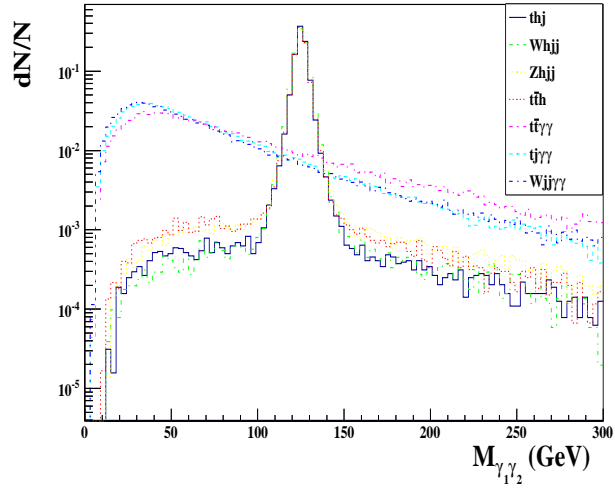}\vspace{-0.5cm}
\caption{Normalized invariant mass distribution of two photons at 14 TeV LHC.}
\label{higgs}
\end{figure}
In Fig.\ref{higgs}, we present the normalized invariant mass distribution of two photons at 14 TeV LHC. Although the $\gamma\gamma$ decay channel has a small branching ratio, it has the advantage of the good resolution on the $\gamma\gamma$ resonance and is also free from the large QCD backgrounds. From Fig.\ref{higgs}, we can see that the spreading of the $\gamma\gamma$ invariant-mass peak at $m_h$ for the signal and the resonant backgrounds is relatively small. We will use a narrow invariant mass window $|M_{\gamma\gamma}-M_h|<5$ GeV to further reduce the non-resonant backgrounds.

\begin{figure}[ht]
\centering
\includegraphics[width=3.5in,height=2.8in]{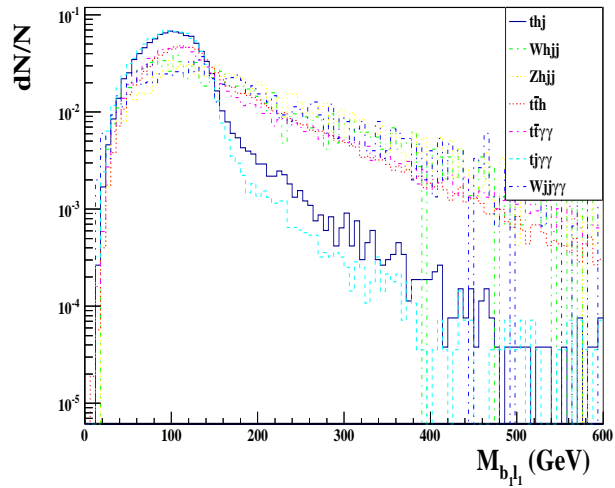}\vspace{-0.5cm}
\caption{Normalized invariant mass distribution of the $b$ jet and lepton at 14 TeV LHC.}
\label{top}
\end{figure}
In Fig.\ref{top}, we plot the normalized invariant mass distribution of the $b$ jet and lepton at 14 TeV LHC, which is another effective cut to remove the backgrounds. From Fig.\ref{top}, we can see that the invariant mass $M_{b_1 \ell_1}$ of the signal is always less than the top quark mass since the leading $b$ jet and lepton in our signal come from the same top quark decay. The same feature also appears in the non-resonant background $tj\gamma\gamma$. But other backgrounds can have higher invariant mass $M_{b_1 \ell_1}$ than the signal. Very similar to $M_{b_1 \ell_1}$, the invariant mass distribution of the diphoton and leading light jet $M_{\gamma_1\gamma_2j}$ also has a peak around the top quark mass in the signal other than the backgrounds, which can be used to further remove the backgrounds.

\begin{table}[ch!]
\fontsize{10pt}{8pt}\selectfont
\begin{center}
{\renewcommand{\arraystretch}{1.25}
\caption{Cut flow of the cross sections for the backgrounds and the signals in the Case-$(I)$, $(II)$ and $(III)$ at 14 TeV LHC, where $\kappa_{tqh}$, $\kappa_{tuh}$ and $\kappa_{tch}$ are assumed to be 0.1 respectively and the symbol $''-''$ stands for the events number less than one. As a comparison, the corresponding results of the resonant production $pp \to t\bar{t} \to thj$ for each case are also listed in the table. \label{cutflow}}
\begin{tabular}{ c c c|  c cc | c  c c c c c}
\cline{1-6}
\hline
&\multicolumn{2}{c|}{\multirow{3}{*}{Cuts}}&\multicolumn{6}{c}{Cross sections (10$^{-3}$ fb$^{-1}$)} \\\cline{4-11}
&&&\multicolumn{3}{c|}{$thj(t\bar{t})$} &\multirow{2}{*}{$t\bar{t}h$ }&\multirow{2}{*}{$Vhjj$}&\multirow{2}{*}{$t\bar{t}\gamma\gamma$}&\multirow{2}{*}{$tj\gamma\gamma$} &\multirow{2}{*}{$Wjj\gamma\gamma$} \\\cline{4-6}
&&&{Case-$I$} &{Case-$II$} &{Case-$III$}      \\\cline{1-11}
\multirow{4}{*}{(1)} &$\Delta R_{ij} > 0.4$,  & $i,j = b,j, \gamma \ \text{or}\  \ell$ & \multirow{4}{*}{2.51(2.32)}&\multirow{4}{*}{1.35(1.16)} &\multirow{4}{*}{1.16(1.16)}
& \multirow{4}{*}{0.035}& \multirow{4}{*}{0.08}& \multirow{4}{*}{4.05}& \multirow{4}{*}{2.92}& \multirow{4}{*}{2.13}\\
&   $p_{T}^b > 25 \ \text{GeV}, $     &$|\eta_b|<2.5$&&&&&\\
&     $p_{T}^\ell > 20 \ \text{GeV}$,    &$|\eta_\ell|<2.0$&&&&&\\
&    $ p_{T}^j > 25 \ \text{GeV}$,    &$|\eta_j|<2.5$&&&&&\\    \hline
(2)& \multicolumn{2}{ c| }{ $p^{\gamma_1}_T>50$ GeV, $p^{\gamma_1}_T>25$ GeV } &2.27(2.10)&1.22(1.05)&1.05(1.05)&0.032&0.007&1.91&1.50&1.28 \\ \hline
(3)& \multicolumn{2}{ c| }{ $M_{b_1\ell_1}<200$ GeV } &2.27(2.10)&1.22(1.05)&1.05(1.05)&0.030&0.005&1.77&1.48&0.85 \\ \hline
(4)& \multicolumn{2}{c|}{$ |M_{\gamma_1\gamma_2}-m_h|< 5$ GeV}&1.99(1.86)&1.06(0.93)&0.93(0.93)&0.022&0.004&0.07&0.09&-\\ \hline
(5)& \multicolumn{2}{c|}{$ M_{\gamma_1\gamma_2j_1} < 300$ GeV}&1.54(1.46)&0.81(0.73)&0.73(0.73)&0.01&0.002&-&0.07&-\\ \hline
\end{tabular}}
\end{center}
\end{table}

According to the above analysis, events are selected to satisfy the following criteria:
\begin{itemize}
  \item exact one isolated lepton with $p_T(\ell_1) > 20$ GeV and $|\eta_{\ell_1}| < 2$.
  \item a hard jet with $p_T(j_1) > 25$ GeV and $|\eta_{j_1}| < 2.5$ and one $b$-jet with $p_T(b_1) > 25$ GeV and $|\eta_{b_1}| < 2.5$;
  \item two photons with $p^{\gamma_1}_{T}>50$ GeV and $p^{\gamma_2}_{T}>25$ GeV and their invariant mass $M_{\gamma_1\gamma_2}$ in the range of $M_h\pm5$ GeV;
  \item the invariant mass of $b$-jet and lepton $M_{b\ell}<200$ GeV£»
  \item the invariant mass of diphoton and leading jet $M_{\gamma_1\gamma_2 j_1} <300$ GeV.
\end{itemize}

In Table \ref{cutflow}, we give the cross sections of the signals in the Case-$(I)$, $(II)$ and $(III)$ and backgrounds after the cut flow at 14 TeV LHC, where $\kappa_{tqh}$, $\kappa_{tuh}$ and $\kappa_{tch}$ are assumed to be 0.1 respectively. From Table \ref{cutflow}, we can see that all the non-resonant backgrounds after the cuts of the two photons are reduced by half while the signals and the resonant backgrounds are hurt slightly. Then, we impose the invariant mass cut $M_{b_1\ell_1}<200$ GeV to remove the backgrounds that do not involve the top quark. Since the photon final states have a good energy resolution in the detector, we require that $M_{\gamma_1\gamma_2}$ be in the range of $120 {\rm GeV} < M_{\gamma_1\gamma_2} < 130 {\rm GeV}$ and $M_{\gamma_1\gamma_2 j_1}<300$ GeV, which can further suppress the backgrounds by half. So at the end of the cut flow, the largest background is $tj\gamma\gamma$, which is followed by $t\bar{t}h$.

\begin{figure}[ht]
\centering
\includegraphics[width=6in,height=2.5in]{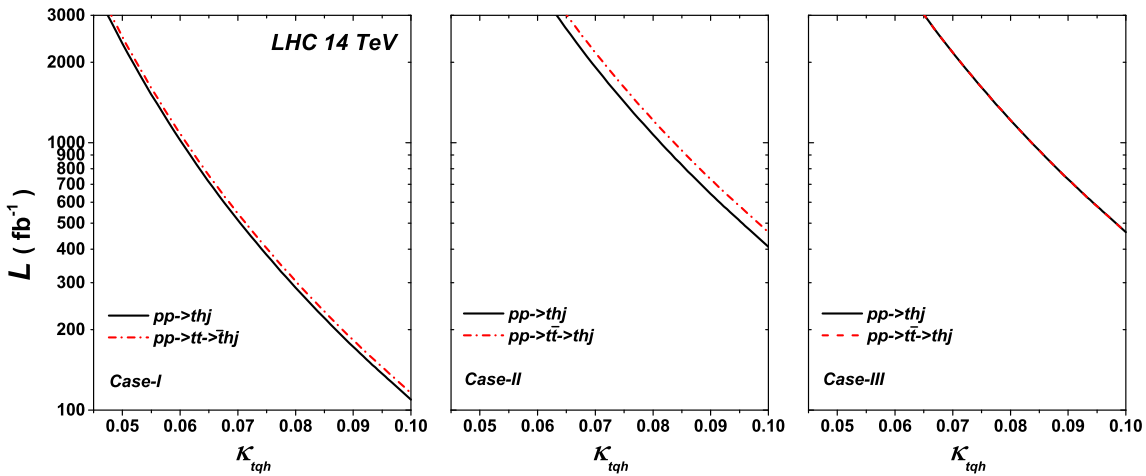}\vspace{-0.5cm}
\caption{Contour plots in ${\cal L}-\kappa_{tqh}$ plane for statistical significance $S/\sqrt{B}=3\sigma$ of $pp \to thj$ at 14 TeV LHC. The conjugate processes have been included in the calculations. The cross section of $t\bar{t}$ is normalized to the approximately next-to-next-to-leading order value $\sigma_{t\bar{t}}=920$ pb \cite{ttbar}. As a comparison, the corresponding results of the resonant production $pp \to t\bar{t} \to thj$ for each case are also displayed.}
\label{ss}
\end{figure}
In Fig.\ref{ss}, we plot the contours of statistical significance $S/\sqrt{B}=3\sigma$ of $pp \to thj$ at 14 TeV LHC for the Case-$(I)$, $(II)$ and $(III)$ in the plane of ${\cal L}-\kappa_{tqh}$. From Fig.\ref{ss}, we can see that the flavor changing couplings $\kappa_{tqh}$ can be probed to 0.047, 0.063 and 0.065 at $3\sigma$ statistical sensitivity by fully calculating the production of $thj$ for the case $(I)$, $(II)$ and $(III)$ respectively, which correspond to the branching ratios $Br(t\to qh)=0.12\%$, $Br(t\to uh)=0.23\%$ and $Br(t\to ch)=0.26\%$ at 14 TeV LHC with ${\cal L} =3000$ fb$^{-1}$. Besides, the corresponding results of the resonant production $pp \to t\bar{t} \to thj$ for each case are also displayed. We can see that the LHC sensitivity to the coupling $\kappa_{tqh}$ from the full calculation of $thj$ production in the Case-$(II)$ can be improved by about 4\% as a comparison with the resonant production $pp \to t\bar{t} \to thj$, while for other two cases, the enhancement is negligible small. Here it should be mentioned that we normalize the leading order cross section of $t\bar{t}$ to the approximately next-to-next-to-leading order value $\sigma_{t\bar{t}}=920$ pb \cite{ttbar}. But the contribution of $qg \to thj$ is calculated at the leading order due to lack of the high order correction. So if assuming that the $k$ factor of the process $qg \to thj$ be the same as $t\bar{t}$, we can expect the sensitivity to the coupling $\kappa_{tqh}$ from full calculation in Case-$(I)$ and $(II)$ will be further increased. Compared with other decay modes of the Higgs boson, our result is close to that of multi-leptons channel in $t\bar{t} \to th(\to WW^*, \tau^+\tau^-, ZZ^*)j$ production \cite{multilepton}, based native scaling in cross section and luminosity at 14 TeV LHC. Although the decay of $h \to b\bar{b}$ has a larger branching ratio and seems more promising \cite{th-6}, the analysis was performed at the parton level without including the parton shower and detector effects. However, these effects are important for the Higgs mass reconstruction and can severely reduce the cut efficiency of Higgs mass window in $b\bar{b}$ channel.

\section{CONCLUSION}
In the work, we investigated the process $pp \to thj$ induced by the top-Higgs FCNC couplings at the LHC. We found that the cross section of $pp \to thj$ can be sizably enhanced in contrast with the SM predictions at 8 and 14 TeV LHC under the current constraints. We studied the observability of top-Higgs FCNC couplings through the process $pp \to t(\to b\ell^{+} \nu_{\ell}) h( \to \gamma\gamma) j$ by including the resonant and non-resonant $hj$ production at 14 TeV LHC. Compared with the resonant production $pp \to t\bar{t} \to thj$, such a full calculation can increase the LHC $3\sigma$ sensitivity to $Br(t \to qh)$ by 4\% and $Br(t \to uh)$ by 10\% at 14 TeV LHC with ${\cal L} =3000$ fb$^{-1}$ because of the contribution of the non-resonant production $qg \to thj$. Finally, the branching ratios $Br(t\to qh)$, $Br(t \to uh)$ and $Br(t \to ch)$ can be respectively probed to $0.12\%,~0.23\%$ and $~0.26\%$ at $3\sigma$ level at 14 TeV LHC with ${\cal L} =3000$ fb$^{-1}$.

\section*{Acknowledgement}
This work was supported by the Australian Research Council, the National Natural Science Foundation of China (NNSFC) under grants Nos. 11275057, 11305049 and 11405047, by Specialized Research Fund for the Doctoral Program of Higher Education under Grant No.20134104120002.

\end{document}